\documentclass[aps,prd,showpacs,eqsecnum,nofootinbib]{revtex4}

\usepackage{epsfig}
\usepackage{graphicx}
\usepackage{dcolumn}
\usepackage{amsmath}
\usepackage{enumerate}
\usepackage{epstopdf} 

\begin{document}

\title{
Inert-Sterile Neutrino: Cold or Warm Dark Matter Candidate
}

\author{
\mbox{Graciela B. Gelmini$^{1}$,} 
\mbox{Efunwande Osoba$^{1}$} and
\mbox{Sergio Palomares-Ruiz$^{2}$}
}

\affiliation{
\mbox{$^1$  Department of Physics and Astronomy, UCLA,
 475 Portola Plaza, Los Angeles, CA 90095, USA}
 \mbox{$^2$ Centro de Fisica Teorica de Particulas (CFTP), Instituto
   Superior Tecnico, P-1049-001, Lisboa, Portugal} 
\\
{\tt gelmini@physics.ucla.edu},
{\tt eosoba@physics.ucla.edu},
{\tt sergio.palomares.ruiz@ist.utl.pt} 
}


\vspace{6mm}
\renewcommand{\thefootnote}{\arabic{footnote}}
\setcounter{footnote}{0}
\setcounter{section}{1}
\setcounter{equation}{0}
\renewcommand{\theequation}{\arabic{equation}}

\begin{abstract} \noindent
In usual particle models, sterile neutrinos can account for the dark
matter of the Universe only if they have masses in the keV range and
are warm dark matter. Stringent cosmological and astrophysical bounds,
in particular imposed by X-ray observations, apply to them. We
point out that in a particular variation of the inert doublet model,
sterile neutrinos can account for the dark matter in the Universe and
may be either cold or warm dark matter candidates, even for masses
much larger than the keV range. These Inert-Sterile neutrinos,
produced non-thermally in the early Universe, would be stable and have
very small couplings to Standard Model particles, rendering very
difficult their detection in either direct or indirect dark matter
searches. They could be, in principle, revealed in colliders by
discovering other particles in the model.   
\end{abstract} 

\pacs{95.35.+d, 98.80.Cq. 12.60.Jv, 14.80.Ly
\hfill CFTP/09-39}
\maketitle 

Since the first indications of the existence of dark matter more than
seven decades ago~\cite{Zwicky}, many different strong pieces of
evidence in its favor have been accumulated (for reviews, see
eg. Refs.~\cite{Jungman:1995df,Bergstrom:2000pn,Bertone:2004pz}). The
presence of dark matter has been revealed so far through its
gravitational effects. Much effort is being devoted to the detection
of dark matter annihilation or decay products or the scattering of
dark matter particles off nuclei. However dark matter may consist of
particles which will not be revealed (at least in the near future) in
this type of searches. We provide here an example of a dark matter
candidate found in a simple extension of the Standard Model (SM),
whose nature could be indirectly proven only through the discovery and
study in colliders of other non-standard particles predicted within
the model. The dark matter particle candidate we study here is a
sterile neutrino with mass in the tens of keV to the tens of GeV range
and produced non-thermally in the early Universe, which can be either
warm dark matter (WDM) or cold dark matter (CDM). 

One or more gauge singlet right-handed (sterile) neutrinos are
included in simple extensions of the SM which can easily accommodate 
neutrino oscillation
data~\cite{numodels,Dodelson:1993je,WDMold,NuSM}. These data show that
at least two of the active neutrinos have a non-zero mass. In many
models sterile neutrinos are the right-handed Dirac mass partners of
the active neutrinos. In some see-saw-inspired models, sterile
neutrinos have large Majorana masses, which leads to three light
(mostly active) neutrinos and several heavier (mostly sterile)
neutrinos, the lightest of which is an attractive dark matter
candidate~\cite{NuSM}. Since this candidate necessarily decays into a
light neutrino and a photon, to constitute the dark matter its
lifetime must be much longer that the age of the Universe. Thus, this
dark matter candidate might be detected through the photons produced
in its decay in  the dark halos of galaxies. Moreover, to have the
required dark matter relic density, the lightest sterile neutrino must
usually have a mass in the keV range, although this depends on the
mechanism through which sterile neutrinos are produced in the early
Universe.    

Relic sterile neutrinos with only standard model interactions are
produced in the early Universe through active-sterile neutrino
oscillations. Sterile neutrinos produced through non-resonant
oscillations~\cite{Dodelson:1993je,WDMold,NuSM} must have masses $M_s$
in the keV range to account for the whole of the dark matter and are
WDM. Through a combination of X-ray and structure formation
constraints, an upper bound $M_s \leq 3-4 $ keV has been
obtained~\cite{WDMold,Xrays,NuSM,Boyarsky:2009ix,Kusenko:2009up} (see,
however, Ref.~\cite{Loewenstein:2009cm} for a very recent weak hint of a
possible signal). Lyman-$\alpha$ forest data has been used to impose
the lower bound of $M_s \geq 5.6 $ keV~\cite{VielWDM1} (see also
Refs.~\cite{SeljakWDM,VielWDM2} for previous bounds) on non-resonantly 
produced sterile neutrinos, or the revised limit of $M_s \geq 8 $ keV
obtained by a new analysis~\cite{Boyarsky:2008xj}, which combined with
the previous upper bounds would exclude non-resonantly produced dark
matter sterile neutrinos. Even disregarding the controversial
Lyman-$\alpha$ bounds, the mass range allowed for these neutrinos is
very restricted because there is an independent lower bound $M_s \geq
1.8 $ keV~\cite{Gorbunov:2008ka,Boyarsky:2008ju} derived from the
analysis of phase space density evolution of dwarf spheroidal
galaxies. In general, these bounds do not consider the possibility of a
very large lepton asymmetry. In the presence of a large lepton asymmetry
$\mathcal{L}\equiv (n_{\nu_e}-n_{\bar{\nu}_e})/s > 10^{-6}$, where
$n_{\nu_e}$ and $n_{\bar{\nu}_e}$ are the number densities of
neutrinos and antineutrinos and $s$ is the entropy density in the
Universe, sterile neutrinos may be produced in the early Universe
through resonant oscillations~\cite{Shi:1998km,Laine:2008pg}.
Considering the upper limit of the lepton asymmetry imposed by Big
Bang Nucleosynthesis (BBN), $\mathcal{L} < 2.5
\times10^{-3}$~\cite{Laine:2008pg}, the range 1~keV $\leq M_s\leq$~50 
keV is in principle allowed for sterile neutrino dark
matter~\cite{Boyarsky:2008ju,Laine:2008pg,Boyarsky:2008mt}. In
slightly more complicated models, sterile neutrino dark matter may be
produced as decay products of, for example, a heavy singlet
scalar~\cite{Shaposhnikov:2006xi,KusenkoPetraki}, or may not
completely thermalize as in low reheating temperature
scenarios~\cite{lowreheat}. Yet, in all these models the X-ray
constraints are important.  
 
Here, we consider a small variation of the SM in which the lightest
sterile neutrino is stable (hence  it does not produce photons as decay
products) and may constitute all of the dark matter. We study a variation
of the Inert Doublet Model~\cite{Ma:2006km,Barbieri:2006dq} (in itself
an extension of the model in Ref.~\cite{Deshpande:1977rw}). In this
model one scalar doublet, $\eta= (\eta^+, \eta_0)$ and three sterile
neutrinos, which we call Inert-Sterile neutrinos, $N_i$ with
$i=1,2,3,$ odd under a new parity $Z_2$, are added to the SM. All the 
particles in the SM are even under the additional $Z_2$ symmetry.
These assignments make the new particles ``inert" because their
couplings to the SM particles are very limited. The leptonic Yukawa
couplings in this model are
\begin{equation}
{\cal L}_Y = f_{ij} (\phi^- \nu_i + \bar \phi^0 l_i) l^c_j + h_{ij} 
(\nu_i \eta^0 - l_j \eta^+) N_j + h.c.~.
\label{yukawa}
\end{equation}
Here $\phi= (\phi^+,\phi^0)$ is the SM scalar doublet field, and $L=
(\nu_i,l_i)$ are the SM lepton fields. Under the extended electroweak
symmetry $SU(2)_L  \times U(1)_Y \times Z_2$, the fields $\eta$, $N$,
$\phi$ and $L$ are in the  (2,1/2;-), (1,0;-), (2,1/2;+)
and (2,-1/2;+) representations respectively. The inert and the
standard doublet scalar also couple through the scalar
potential~\cite{Barbieri:2006dq},   
\begin{equation} 
V = \mu_1^{2} \Phi^\dagger \Phi + \mu_2^{2} \eta^\dagger \eta + 
\lambda_1 (\Phi^\dagger \Phi)^{2} + \lambda_2 
(\eta^\dagger \eta)^{2} + \lambda_3 (\Phi^\dagger \Phi)(\eta^\dagger
\eta) + \lambda_4 (\Phi^\dagger \eta)(\eta^\dagger \Phi) + 
\frac{1}{2} \lambda_5 [(\Phi^\dagger \eta)^{2} + h.c.]~.
\label{potential}
\end{equation} 
In particular the last quartic coupling provides the mass splitting
between the two physical inert neutral scalar particles $\eta_H=\sqrt
2~ \rm{Im}(\eta^{0})$ and $\eta_L = \sqrt
2~\rm{Re}(\eta^{0})$~\cite{Ma:2006km, Barbieri:2006dq}, which  
are the heaviest and the lightest for $\lambda_5<0$ (otherwise the two
would be exchanged) 
 \begin{equation}
 m_{\eta_H}^2 -m_{\eta_L}^2 =  |\lambda_5| v^2~.
 \label{eta-mass-split}
\end{equation}
The masses of the inert scalar bosons are,
\begin{eqnarray} 
m_{\eta^+}^2 &=& \mu_2^2 + \lambda_3 v^2/2\cr 
m_{\eta_H}^2 &=& \mu_2^2 +  (\lambda_3 + \lambda_4 - \lambda_5) v^2/2 =
\mu_2^2 +  (\lambda_L - 2 \lambda_5) v^2/2 \cr 
m_{\eta_L}^2 &=& \mu_2^2 +  (\lambda_3 + \lambda_4 + \lambda_5) v^2/2 =
\mu_2^2 +  \lambda_L  v^2/2~.  
\label{masses} 
\end{eqnarray} 
Here $v/\sqrt {2} = 174$ GeV is the vacuum expectation value (VEV) of
the SM Higgs field (the inert scalar does not acquire a VEV),
$m_{\eta^\pm}$ is the mass of the charged scalars, $\lambda_5$ has
been chosen to be real and we define $\lambda_L = \lambda_3 +
\lambda_4 + \lambda_5$. The only other terms in the Lagrangian allowed
by the $Z_2$ symmetry are Majorana mass terms for the Inert-Sterile
neutrinos, 
\begin{equation}
\frac{1}{2} M_{i} N_i N_i + h.c.~.
\label{Majorana-mass}
\end{equation}
The $Z_2$ symmetry forbids sterile-active neutrino mixings. The
$N_i$'s are not the Dirac mass partner of the $\nu_i$ as in usual
extensions of the SM and active neutrino Majorana masses are generated
at one-loop level. The active neutrino mass matrix elements
are~\cite{Ma:2006km}
\begin{equation}
({\cal M}_\nu)_{ij} = 
\sum_k {h_{ik} h_{jk} \frac{M_k}{16 \pi^{2}} \left[
    \frac{m_{\eta_H}^2}{m_{\eta_H}^2 -M_k^2}
    \ln \left(\frac{m_{\eta_H}^2}{M_k^2} \right) -
    \frac{m_{\eta_L}^2}{m_{\eta_L}^2  -M_k^2} 
\ln \left({\frac{m_{\eta_L}^2}{M_k^2}}\right)\right]}~.
\label{active-mass-1}
\end{equation}
We will assume in what follows that $m_{\eta_H}$ is of the order of
100 GeV and $m_{\eta_L}$  of the order of tens of GeV, thus  the first
term in Eq.~\ref{active-mass-1} is dominant.  

The $Z_2$ parity implies that the lightest inert particle is stable
and thus a good dark matter candidate. Both the lightest inert
scalar~\cite{Ma:2006km,Barbieri:2006dq,Boehm:2006mi,LopezHonorez:2006gr,
TytgatInert,Gustafsson:2007pc,Dolle:2009fn,DolleInert}
and the lightest sterile
neutrino~\cite{Ma:2006km,Kubo:2006yx,Sierra:2008wj} could be dark matter
candidates. We will assume the second possibility.
 
In Refs.~\cite{Kubo:2006yx,Sierra:2008wj} it was assumed that the mass
difference between $\eta_L$ and $\eta_H$ is small, i.e. the coupling
$\lambda_5$ is  very small. In this case, in order to generate the
observed active neutrino masses, the $h_{ij}$ couplings cannot be very
small. In addition, it was assumed that $m_0= (m_{\eta_H}^{2} +
m_{\eta_L}^{2})/2 > M_1, M_2, M_3$ and the lightest $N_i$ is produced
thermally. Under these assumptions, Ref.~\cite{Kubo:2006yx} found that 
the lightest Inert-Sterile neutrino can be CDM and  account for the
whole of the dark matter if its mass is in the range 7~GeV to 300~GeV. 
 
Here we will explore a range of values of the coupling constants
different from those previously considered, namely $\lambda_5$ not very
small and $h_{ij}$ Yukawa couplings small enough to ensure that the
sterile neutrinos $N_i$ are never in equilibrium in the early
Universe. We will not study the flavor structure of the couplings
$h_{ij}$, but only their order of magnitude. We call generically
$h_1$, $h_2$, $h_3$  the couplings of $N_1$, $N_2$ and $N_3 $,
respectively. We assume a hierarchy in the couplings, with
$h_1<h_2\simeq h_3$. We also assume that only the lightest sterile 
neutrino, which we take to be $N_1$, is lighter than the lightest
inert scalar $\eta_L$ and hence, it is the dark matter candidate. The
$\eta_L$ particles are produced thermally in the early Universe and
decouple when they are non-relativistic. The subsequent late decay of
the $\eta_L$ produces the Inert-Sterile $N_1$ relic particles that 
now constitute the dark matter. In this scenario, depending on the
mass, abundance and lifetime of $\eta_L$, the  $N_1$ can be either CDM
or WDM and account for the whole of the dark matter with mass in the range
$\sim$few keV to tens of GeV. We will show that all requirements
on the model are fulfilled: active neutrino masses of the right order
of magnitude are obtained, the upper bound on the $h_{i,j}$ from $\mu
\to e \gamma$ is easy to fulfill, all $N_i$ producing reactions in the
early Universe are out of equilibrium and the necessary relic density
and decay rate of $\eta_L$ for different values of the $\eta_L$ and
$N_1$ masses are obtained, while respecting all the collider  and
other bounds imposed on the model.

Let us see first how large the Yukawa couplings $h$ must be to get
reasonable values for the active neutrino masses,
i.e. $\left(M_\nu\right)_{i,j}\simeq 10^{-1}$ eV. Using
Eq.~\ref{active-mass-1}, and assuming that $\eta_L$ is significantly 
lighter than $\eta_H$, that $M_{2,3}$ is of the same order of
magnitude but larger than $m_{\eta_H}$ and that the contributions of
$N_2$ and $N_3$  are dominant, we get 
\begin{equation}
h_{2,3} \simeq 0.7 \times 10^{-5} \, \left(\frac{M_{2,3}}{100 \, {\rm
    GeV}} \right)^{1/2} \, \left(\frac{100 \, {\rm GeV}}{m_{\eta_H }}
\right) \, \left[\ln{ \left(\frac {M_{2,3}^2}{m_{\eta_H}^2}
    \right)}\right]^{-1/2}~.
\label{neutrinomasshbound-1}
\end{equation}
Notice that when $m_{\eta_H}$ is large with respect to $m_{\eta_L}$,
Eq.~\ref{eta-mass-split} implies that $m_{\eta_H} \simeq
\sqrt{|\lambda_5|} \, v$. Moreover, when  $M_{2,3}$ are larger than but 
similar to $m_{\eta_H}$, the logarithm in
Eq.~\ref{neutrinomasshbound-1} is close to 1, thus
\begin{equation}
h_{2,3} \simeq  \frac{3 \times 10^{-6}}{\sqrt{|\lambda_5|}} \left(\frac
{M_{2,3}} {100 \, \rm{GeV}} \right)^{1/2}~. 
\label{neutrinomasshbound-2}
\end{equation}

\begin{figure}[t]
\includegraphics[width=0.4\textwidth,clip=true,angle=0]{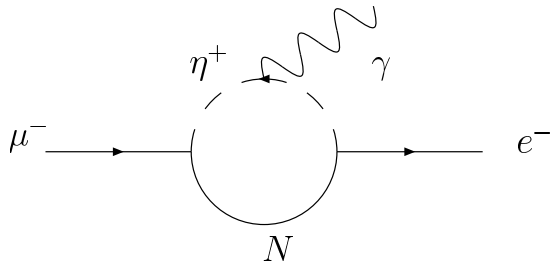}
\caption{Diagram for $\mu \to e\gamma$ transition in the variation of
  the inert doublet model we consider here.} 
\label{ratio-0}
\end{figure}

Lepton flavor transitions like the $\mu \to e\gamma$ process in
Fig.~\ref{ratio-0} occur in this model. The branching ratio,
$B_{\mu\to e\gamma} = \Gamma_{\mu \to e\gamma}/\Gamma_{\mu \to
  e\nu\nu}$ in the inert doublet model is~\cite{Ma:2001mr,Kubo:2006yx}
\begin{eqnarray}
B_{\mu\to e\gamma}&=&\frac{192 \, \pi^3 \, \alpha}{G_F^2} \,
\left(\left|\sum_j \frac{h_{\mu j} \, h_{ej}}{4 \, (4\pi)^2 \,
  m_{\eta^-}^2} \, F_2 \,
\left(\frac{M_j^2}{m_{\eta^-}^2}\right)\right|\right)^2,   
\end{eqnarray}
where $\alpha$ is the fine structure constant and $G_F$  is the Fermi
constant. For  $M_{2,3} \simeq m_{\eta^\pm}$ the function $F_2 (x) =
[1-6x+3x^2+2x^3-6x^2\ln(x)] [6(1-x)^4]^{-1}$ is 
$F_2(1) \simeq 1/12$, whereas for $M_1 < m_{\eta^\pm}$ it is $F_2(0)
\simeq 1/6$~\cite{Ma:2001mr}. The experimental upper bound on the
branching ratio, $B(\mu \to e\gamma) \leq 1.2 \times
10^{-11}$~\cite{Brooks:1999pu}, implies    
\begin{equation}
h_{2,3}\leq 2 \times10^{-2} \left(\frac{m_{\eta^\pm}}{100 \,
  \rm{GeV}}\right)
\end{equation}
for $h_1 << h_{2,3}$.

\begin{figure}[t]
\includegraphics[width=0.4\textwidth,clip=true,angle=0]{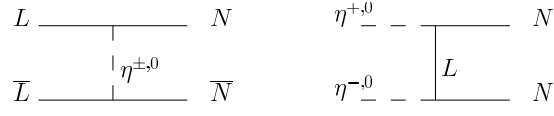}
\caption{Production processes of Inert-Sterile neutrinos in the early
  Universe.} 
\label{N-production}
\end{figure}

Let us now see how small the couplings $h_{ij}$ must be in order for
the $N_i$ to never be in equilibrium in the early Universe. The upper
bounds are particularly important for $N_2$ and $N_3$, whose generic
couplings, $h_2$ and $h_3$, are larger than the coupling $h_1$ of
$N_1$. The $N_i$ can be produced through the reactions in
Fig.~\ref{N-production}, i.e. two to two reactions $L \bar L \to N_i
N_i$ mediated by any of the four physical inert particles $\eta_H,
\eta_L, \eta^+ , \eta^-$, which here we call now generically $\eta$,
or $\eta \eta \to N_i N_i$ mediated by $L$. The $N_i$ could in
principle be produced through the decay $\eta \to N_i L$ and the
inverse decay of $ \eta L\to N_i$. The production rate for $N_2$, for
example, is
\begin{eqnarray}
\Gamma_{ N_2} &=& \sum_{L} \left(2<\sigma v>_{L\bar{L}  \to  N_2 N_2}
+ <\sigma v>_{L\bar{L} \to  N_2 N_3} \right)n_L^2/n^{eq}_{N_2}
\nonumber  \\  
&+&\sum_{\eta} \left(2<\sigma v>_{\eta \eta  \to  N_2 N_2} + <\sigma
v>_{\eta \eta \to  N_2 N_3} \right)n_\eta^2/n^{eq}_{N_2} 
\label{Gamma}
\end{eqnarray}
where $n^{eq}_{N_2}$ is the $N_2$ equilibrium number density which
appears in the equation as a normalization factor, and $n_L$ and
$n_\eta$ are the  number densities of the standard  leptons and the
inert scalars respectively, at the temperature considered.

Eq.~\ref{Gamma} is derived from the Boltzmann equation for the
production of $N_i$  ($i=1,2,3$) in the process $a b \to N_i c$, where
$a,b$ and $c$ are particles and we assume that only the initial
particles, $a$ and $b$, have initially a non zero particle density. If
the  particles $a$ and $b$ are in equilibrium, assuming
Maxwell-Botzmann density distributions, the time evolution of the
number density $n_{N_i}$ depends on the number densities of particles
$a$ and $b$ in the following manner (see e.g. Eqs. (5.8) and (5.23) of
Chap. 5 of Ref.~\cite{Kolb-Turner}) 
\begin{equation}
\frac{d n_{N_i}} {dt} +3H n_{N_i} = \sum_{a,b,c}  n_a  n_b <\sigma_{ab
  \to N_i c} |v|> .
\end{equation}
As usual, it is convenient to change variables to $Y \equiv n_{N_i}/s$
and $x \equiv m_{N_i}/T$ (see Eq. (5.16) of Ref.~\cite{Kolb-Turner})
and obtain  
\begin{equation}
\frac{d Y} {dx} = \frac{1}{H \, x \, s} \sum_{a,b,c}  n_a \, n_b \,
<\sigma_{ab \to N_i c} |v|>
\label{Y}  
\end{equation}
Now, dividing and multiplying the right-hand side of Eq.~\ref{Y} by
$n_{N_i}^{eq}$ as a normalizing function, one gets 
\begin{equation}
\frac {x}{Y^{eq}} \frac{d Y} {dx} = \frac{\Gamma_{N_i}}{ H} 
\label{Y}
\end{equation}
where $\Gamma_{N_i}$ is defined as in Eq.~\ref{Gamma}
above. Eq.~\ref{Y} is equivalent in this case to Eq. (5.26) of
Ref.~\cite{Kolb-Turner} and shows that $Y_{N_i}$ is never
significantly different from zero if $\Gamma_{N_i}/H < 1$.

Following Refs.~\cite{Srednicki:1988ce,Gondolo:1990dk} and making use of
Refs.~\cite{Cheung:2004xm,Boehm:2003hm}, for relativistic $\eta$,
$N_i$ and $L$ the thermal averaged cross section of $L\bar{L}\to N_i
N_i$ and of $\eta\eta \to N_i N_i$ are approximately given by
\begin{eqnarray}
 & < \sigma v>_{L\overline{L}\to N_i N_i}&\simeq 0.7 \times
    10^{-1} \, \frac{h_i^4}{T^2}~,\\ 
 &< \sigma v>_{\eta\eta \to N_i N_i}& \simeq 3 \times 10^{-1} \,
    \frac{h_i^4}{T^2} \, \ln \left(\frac{4 \,
      T^2}{M_i^2+m_{\eta}^2}\right),    
\end{eqnarray}
which show that the process $\eta \eta \to  N_i N_i$ is dominant and
\begin{equation}
\Gamma_{N_i} \simeq \Gamma_{\eta \eta \to  N_i N_i} \simeq
0.7 \times 10^{-1} \,  h_i^4 \, T \, \ln \left(\frac{4 \,
  T^2}{M_i^2+m_{\eta}^2}\right)~.  
\end{equation}
The production is out of equilibrium if the rate is smaller than
the expansion rate of the Universe, $H$,  
\begin{equation}
\Gamma_{N_i}< H = 1.66 \, \sqrt{g_*} \, \frac{T^2}{M_{\rm{Pl}}}~, 
\label{condition}
\end{equation}
where $g_*$ is the number of degrees of freedom and $M_{\rm Pl}$ is
the Planck mass. Since the right-hand side of Eq.~\ref{condition} decreases faster than the
left-hand side for decreasing $T$, if the condition is fulfilled for
the smallest $T$ value in the range considered, i.e. the smallest $T$
for which all the particles involved in the production are
relativistic, then it is fulfilled for all larger $T$.
   
At high temperatures  $T> M_{2,3} \simeq m_{\eta_H}$ we need to write
the condition in Eq.~\ref{condition} at $T \simeq M_k \simeq
m_{\eta_H}$. Thus, the production of  relativistic $N_{2,3}$ is out of
equilibrium at $T> M_{2,3} \simeq m_{\eta_H}$ if 
\begin{equation}
  h_{2,3}< 2 \times10^{-4}  \, \left(\frac{g_*}{106.75}\right)^{1/8} \,
  \left( \frac {M_{2,3}}{100 \, \rm{GeV}}\right)^{1/4}~. 
\label{equi1-for2and3}
\end{equation}
Since we are assuming  $M_1 < m_{\eta_L} << M_{2,3}$, the condition in
Eq.~\ref{condition} for relativistic $N_1$ and $\eta_L$ must be taken
at $T \simeq m_{\eta_L}$, thus the production of relativistic $N_1$
from relativistic $\eta_L$ is out of equilibrium if 
\begin{equation}
h_1 < 2 \times 10^{-4} \, \left(\frac{g_*}{106.75}\right)^{1/8} \,
\left(\frac{m_{\eta_L}}{10 \, \rm{GeV}}\right)^{1/4}~. 
\label{equi1-for1}
\end{equation}
At temperatures in the range ${M_{2,3}}>T>m_{\eta_L}$, in which the
$N _{2,3}$ are non-relativistic (but $\eta_L$ and $L$ are
relativistic), the relevant thermal average cross sections for 
$N_{2,3}$ production are  approximately
\begin{eqnarray}
<\sigma v>_{L\bar{L}\to N_i N_i} & \simeq& 0.8 \times 10^{-2} \,
\frac{h_i^4}{T^2} \, \exp{\left(-\frac{2 M_i}{T}\right)}~,\\ 
<\sigma v>_{\eta \eta \to N_i N_i}  &\simeq& 2 \times 10^{-1} \,
\frac{h_i^4}{T^2} \, \exp{\left(-\frac{2 M_i}{T}\right)}~. 
 \end{eqnarray}
The production is again dominated by the $\eta_L \eta_L \to N_i N_i$
process, thus 
\begin{equation}
\Gamma_{N_i} \simeq \Gamma_{\eta_L \eta_L \to N_i N_i} \simeq 0.7
\times 10^{-1} \, h_i^4 \, \frac{T^{5/2}}{M_i^{3/2}} \, \exp{\left(-
  \frac{M_i}{T}\right)}~.
\end{equation}
Because this rate decreases faster than $H$, if $\Gamma_{N_i} < H$ is
fulfilled at $T = M_{2,3}$ where $\Gamma$ is maximum within the $T$
interval, the process will be out of equilibrium for lower values of
$T$, thus we obtain   
\begin{equation}
 h_{2,3} < 3 \times 10^{-4} \, \left(\frac{g_*}{106.75}\right)^{1/8}
 \, \left(\frac{M_{2,3}}{100 \, \rm{GeV}}\right)^{1/4}~.   
\label{equi2}
\end{equation}
For still lower temperatures $T < m_{\eta^{\pm,0}} $, for which all
the inert bosons are non relativistic but the $N_1$ are relativistic,
we need to verify that the $N_1$ are not produced thermally (recall we
are assuming that  $m_\eta^{\pm,0} > M_1$). In this case
\begin{equation}
<\sigma v>_{\eta \eta \to N_1 N_1} \simeq  \frac{3}{4 \pi } \, h_1^4 \,
\frac{M_1^2}{ m_{\eta }^4}~, 
\end{equation}
and 
\begin{equation}
\Gamma_{N_1} \simeq \Gamma_{\eta \eta \to N_1N_1} = 10^{-2} \, h_1^4
\, \frac{M_1^2}{m_{\eta}} \, \exp{\left( -\frac{2 \,
    m_{\eta}}{T}\right)}~. 
\end{equation}
This rate decreases faster than $H$ as $T$ decreases, thus if
$\Gamma_{N_1} / H <1$ at the highest temperature in the range 
considered, $T= m_\eta$, the condition is fulfilled at any lower $T$.
Thus,  
\begin{equation}
h_1 < 3 \times10^{-2} \, \left(\frac{g_*}{106.75}\right)^{1/8} \,
\left(\frac{m_{\eta_L}}{10 \, \rm{GeV}}\right)^{3/4}
\left(\frac{\rm{MeV}}{M_1}\right)^{1/2}~.  
\end{equation}
After considering all the required upper bounds on the $h_{ij}$ Yukawa
couplings, we conclude that Eq.~\ref{equi1-for2and3} provides the most
restrictive upper bound on the Yukawa couplings of the heaviest inert
sterile neutrinos, $h_{2,3}$, and they are compatible with the value
assigned to $h_{2,3}$ in Eq.~\ref{neutrinomasshbound-1}, which is
necessary to account for the active neutrino masses. The most
restrictive bound on $h_1$, the Yukawa couplings of the lightest
Inert-Sterile neutrino $N_1$, will be given in Eq.~\ref{h1decaybound}
below and is derived from our requirement of a long enough lifetime of
the lightest inert bosons $\eta_L$ into $N_1$.

Let us now consider the decays of the $\eta^{\pm}$ and $\eta_H$ into
Inert-Sterile neutrinos. If $m_\eta^\pm > m_\eta^0 +m_W$, then the
process $\eta^{\pm} \to \eta^0 + W$ can occur. The branching ratio of
the decay mode $\eta^\pm \to N_i L$ with respect to the dominant
$\eta^{\pm} \to \eta^0 + W$ mode is proportional to the ratio of the
couplings $h^2_i / g_W^2$, where $g_W$ is the weak coupling. Using
the value of $h_{2,3}$ necessary to produce the active neutrino
masses, given in Eq.~\ref{neutrinomasshbound-2} with $|\lambda_5|
\simeq 0.2$ for example, $h_i^2 / g_W^2 \simeq 10^{-10} \,
(M_i/100 \, \rm{GeV})$, which is negligible. Thus, the heavier Inert-Sterile
neutrinos $N_{2,3}$ are not produced in the decays of the inert
charged  bosons. Neither the lightest Inert-Sterile neutrino is
produced in these decays, since $h_1 << h_{2,3}$. If, instead
$m_{\eta^\pm} < m_{\eta^0} +m_W$, the 3-body decay $\eta^{\pm} \to \eta^0
+ L + \bar{L}$ dominates the decay of $\eta^\pm$; the branching ratio 
of $\eta^\pm \to N_i L$ then goes as $h_i^2 / g_W^4 \simeq 10^{-11} \,
(M_i/100 \, {\rm GeV})$ for the heavier Inert-Sterile neutrinos. The
branching ratio is even smaller for $N_1$. Again, the decay of the
charged inert bosons into the Inert-Sterile neutrinos $N_i$ is
negligible. For the decays of the heavier neutral inert boson $\eta_H$
the same arguments apply but changing the $W$'s by $Z$'s. Thus the
Inert-Sterile neutrinos are not produced in the decays of $\eta^{\pm}$
and $\eta_H$.
 
We need to insure that $\eta_L$, the lightest inert scalar particle,
is produced thermally in the early Universe and that it is in 
equilibrium before decoupling while it is already non-relativistic, at
freeze-out,  $T_{f.o.} < m_{\eta_L}$. The dominant processes that
maintain the $\eta_L$ particles in equilibrium depend on the couplings
of $\eta_L$ with the SM particles. The $\eta_L$ gauge couplings and
its couplings in the scalar potential are the same that occur in the
inert doublet model in the absence of sterile neutrinos. Using the
same couplings, in
Ref.~\cite{Barbieri:2006dq,LopezHonorez:2006gr,Dolle:2009fn} $\eta_L$
with mass in the GeV range are found to be good dark matter
candidates. We want instead that the $\eta_L$ decay into the lightest
inert sterile neutrino $N_1$, which constitutes the dark matter
now. After the $\eta_L$ particles decay through the process $\eta_L
\to N_1  \nu_i$, there is one $N_1$ per each $\eta_L$. In order for
$N_1$ to account for the whole of the dark matter, the number density
of  $\eta_L$ at their decoupling must be larger for the case considered
here than in the scenarios in which they constitute the dark 
matter~\cite{Barbieri:2006dq,LopezHonorez:2006gr,Dolle:2009fn}. The
number density $n_{N_1}$  that is needed for non-relativistic $N_1$ to
be the dark matter at present, must be the same relic number density
$n_{\eta_L}$  the $\eta_L $ should have at present had they not
decayed.  Thus the relic density of $N_1$ is now $n_{N_1} M_1=
n_{\eta_L} M_1$ and 
\begin{equation}
\Omega_{N_1} h^2 =\Omega_{\eta_L} h^2 \, \left(\frac {M_1} {m_{\eta_L}}
\right)~,
\label{Omega} 
\end{equation}
where $\Omega_{\eta_L} h^2$ is the relic density the $\eta_L$ would
have at present if they were stable. When the $N_1$ can be either CDM
or WDM we require the  $N_1$ density to be that of the observed
relic density of dark matter $\Omega_{\rm{DM}} h^2 = 0.1099 \pm
0.0062$~\cite{WMAP5}. If the $N_1$ are instead hot dark matter (HDM)
we should impose the upper bound $\Omega_{N_1} h^2 \leq 0.014 \equiv
\Omega_{\rm{HDM-max}} h^2$ (the 95\% CL on the relic density of light
neutrinos)~\cite{WMAP5}.  

\begin{figure}[t]
\includegraphics[width=0.4\textwidth,clip=true,angle=0]{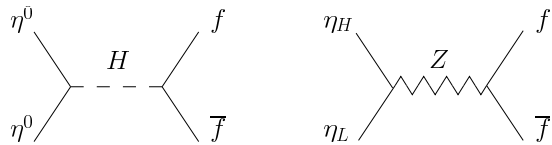}
\caption{Dominant $\eta_L$ annihilation channels into Standard Model fermions $f$.} 
\label{eta-annihilation}
\end{figure}

If $m_{\eta_L} > m_W$, the $\eta_L$ annihilate efficiently into two W
bosons and their relic density is too small even to constitute the
bulk of the dark matter, thus we are not interested in this mass
range. When $m_{\eta_L} < m_W$, the processes in
Fig.~\ref{eta-annihilation} and their inverse processes keep $\eta_L$
in equilibrium. The lightest scalar $\eta_L$ coannihilates with the
heaviest inert scalar partner $\eta_H$. The coannihilation $\eta_H \eta_L \to
f \overline{f} $ into SM fermions $f$ is mediated by the Z boson and its cross section
depends on the mass spittling $\Delta =m_{\eta_H}- m_{\eta_L}$ which
in turn, depends on $\lambda_5$ (see Eq.~\ref{eta-mass-split}). The
$\eta_L$ also coannihilates with $\eta^\pm$, via $W^\pm$ exchange, with
a cross section which depends on the mass split between them. The
process $\eta_L\eta _L\to f\bar{f}$  via Higgs exchange also keeps
$\eta_L$ in equilibrium, and in the particular range of masses we
explore below is the dominant process. We use the public code
MicrOMEGAs~\cite{Belanger:2007zz} to compute the $\eta_L$ relic
density.

The decay $\eta_L \to N_1 L$ must happen after the $\eta_L$ freeze-out
at $T_{f.o.} = m_{\eta_L}/ x_f$ where $x_f $ is in the  20 to
30 range. Thus, the decay rate must be  $\Gamma_{\eta \to N_1 L} \simeq
h_1^2 m_{\eta_L} / 16\pi< H$ for $T>  T_{\rm{decay}} $ and
$\Gamma_{\eta \to N_1 L} \simeq H$ for $T= T_{\rm{decay}}$ with
$T_{\rm{decay}} < T_{f.o.}$. These conditions lead to the most
stringent bound on $h_1$ 
\begin{equation}
h_1< 2 \times 10^{-9} \, \left( \frac{20}{x_f}\right)
\left(\frac{m_{\eta_L}}{10\, \rm{GeV}}\right)^{1/2} \,
\left(\frac{g_*}{10.75}\right)^{1/4}.  
\label{h1decaybound}
\end{equation}
Note that this bound on $h_1$ is consistent with the previous
requirements.

\begin{figure}[t]
\hspace{-0.7cm} 
\includegraphics[width=0.76\textwidth,clip=true,angle=0]{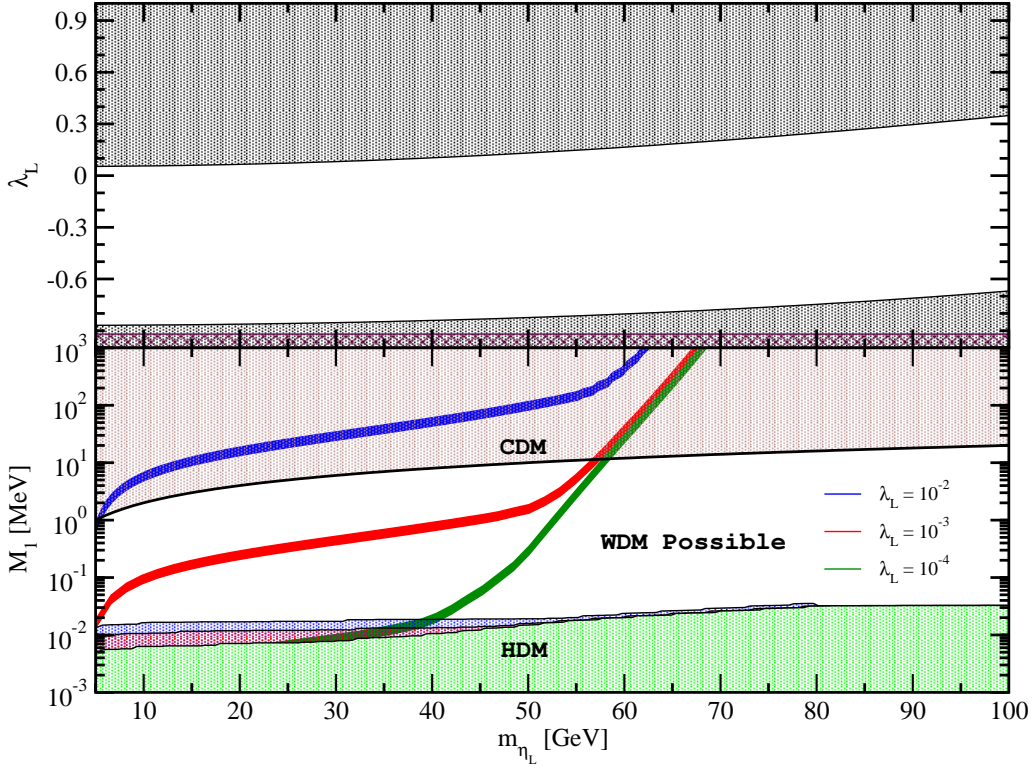}  
\caption{In both panels, $M_{\rm Higgs} = 160$ GeV, $m_{\eta_H} = 125$ GeV
  and $m_{\eta^\pm} = 130$ GeV. {\sl Upper panel:} The shaded areas
  correspond to forbidden values of $\lambda_L$ from vacuum stability
  (cross-hatched violet region) and perturbativity (shaded gray
  region) arguments. {\sl Lower panel:} From top to bottom, the blue,
  red and green colored narrow strips show the regions where $N_1$
  would have the right dark matter density for the corresponding
  values of $\lambda_L$. The unshaded background region corresponds to
  the range in Eq.~\ref{m-WDM-bounds}, where the $N_1$ may constitute
  WDM (above it, $N_1$ can only be CDM and below it, only  HDM). For any
  particular value of $T_{\rm{decay}}$ between 5 MeV (upper boundary
  of unshaded region) and $ m_{\eta_L}/x_f$ (lower boundary of
  unshaded region, which depends on $\lambda_L$ through $x_f$) there
  is one value of $M_1$, given by Eq.~\ref{m-WDM}, within the unshaded
  background region for which $N_1$ would be WDM (and it would be CDM
  for all larger values of $M_1$ and HDM for all smaller ones). In
  order for $N_1$ to be allowed as HDM, its mass must be at least a
  factor of $\Omega_{\rm{DM}} h^2/\Omega_{\rm{HDM-max}} h^2 =
  0.1099/0.014 \simeq 8$ smaller than that corresponding to the center
  of the colored bands for a given $m_{\eta_L}$.}  
\label{relicplot-1}
\end{figure}

We can now show that the Inert-Sterile neutrinos produced in this
model may be either WDM or CDM, which are characterized by the
free-streaming length $\lambda_{fs}$~\cite{Lin:2000hz,Hisano:2000kn} 
\begin{equation}
\lambda _{\rm fs}= 2 \, r \, t_{\rm EQ} \, (1+z_{\rm EQ})^2 \,
\ln\left(\sqrt{1 + \frac{1}{r^2 \, (1+z_{\rm EQ})^2}} + \frac{1}{r \,
  (1+z_{\rm EQ})^2}\right)~. 
\end{equation}
Here the subscript ${\rm EQ}$ denotes matter-radiation equality and
$r=a(t)p(t)/M_1$, where $a(t)$ and $p(t)$ are the scale factor of 
the Universe and the dark matter particle characteristic momentum at
time $t$ respectively. As the Universe expands, the ratio $r$ remains 
constant. At the time of matter-radiation equality, $\lambda_{fs} $
must be $0.1$~Mpc~\cite{Colin:2000av} for WDM, which fixes $r \simeq
10^{-7}$. At the moment of decay of  the $\eta_L$ (we make the 
approximation of instantaneous decays) the scale factor of the
Universe  is $a\simeq T_o/T_{\rm decay}$,  where $T_o$ is the photon
temperature today, and the momentum of the relativistic $N_1$ decay
products is $m_{\eta_L}/2$. Thus, $r \simeq T_o \, m_{\eta_L}/\left(2 \,
T_{\rm{decay}} \, M_1\right)$. Therefore, $r=10^{-7}$ fixes the mass of
$N_1$ to be  
\begin{equation}
\left(M_1\right)_{\rm WDM} \simeq 2.4 \, {\rm MeV} \,
\left(\frac{m_{\eta_L}}{10 \, {\rm GeV}}\right) \, \left(\frac{5 \,
  {\rm MeV}}{T_{\rm decay}}\right)~.  
\label{m-WDM}
\end{equation}
Given a particular $T_{\rm decay}$, Eq.~\ref{m-WDM} provides the $N_1$
mass for which the $N_1$ would constitute WDM. Heavier $N_1$ (smaller
$\lambda _{\rm fs}$) would be CDM and lighter ones (larger $\lambda
_{\rm fs}$) HDM.  

We require the decay temperature to be $T_{\rm decay} \gtrsim 5$~MeV, in
order not to affect the success of BBN predictions, and $T_{\rm decay}
< m_{\eta_L}/ x_f$, because the decays of $\eta_L$ happen after they
decouple. Thus, the range of masses for which $N_1$ could be a good
WDM candidate is 
\begin{equation}
24 \, {\rm keV} \, \left(\frac{x_f}{20}\right) < \left(M_1\right)_{\rm
  WDM} <  2.4 \, {\rm MeV} \, \left(\frac{m_{\eta_L}}{10 \,{\rm
    GeV}}\right)~. 
\label{m-WDM-bounds}
\end{equation}

Finally, in order to choose suitable sets of parameters there are a
number of constraints that need to be considered. The null result for
the process $e^+e^- \to  Z^*\to \eta_H\eta_L$ in LEP II searches for
neutralinos, imposes the bound $m_{\eta_H} >120$~GeV when $m_{\eta_L} <
80$~GeV~\cite{Lundstrom:2008ai}. Alternatively, in a range of
parameters we will not explore, the neutral inert boson mass
difference must be $m_{\eta_H} - m_{\eta_L}<
8$~GeV~\cite{Lundstrom:2008ai} for $m_{\eta_L}+m_{\eta_H} >m_Z$ due to
the LEP I measurement of the Z-width, which implies $m_{\eta_L} >
40$~GeV. In addition, the suitable set of parameters should also be
within the allowed range provided by electroweak precision
measurements~\cite{Barbieri:2006dq,Dolle:2009fn}.    
  
There are also constraints on the $\lambda$ couplings in the scalar
potential, Eq.~\ref{potential}. Vacuum stability of the scalar potential
imposes~\cite{Barbieri:2006dq}  
\begin{eqnarray}
\lambda_{1,2} >0,\lambda_{2} <1 \nonumber \\
\lambda_3,\lambda_L-  \lambda_5- |\lambda_5| >-2\sqrt{\lambda_1
  \lambda_2},  
\label{stability} 
\end{eqnarray} 
and perturbativity  of the scalar potential
imposes~\cite{Barbieri:2006dq}  
\begin{equation}  
{\lambda_3}^2+(\lambda_L-\lambda_5)^2+{\lambda_5}^2< 12 {\lambda_1}^2.
\label{pert}
\end{equation}

\begin{figure}[t]
\includegraphics[width=0.76\textwidth,clip=true,angle=0]{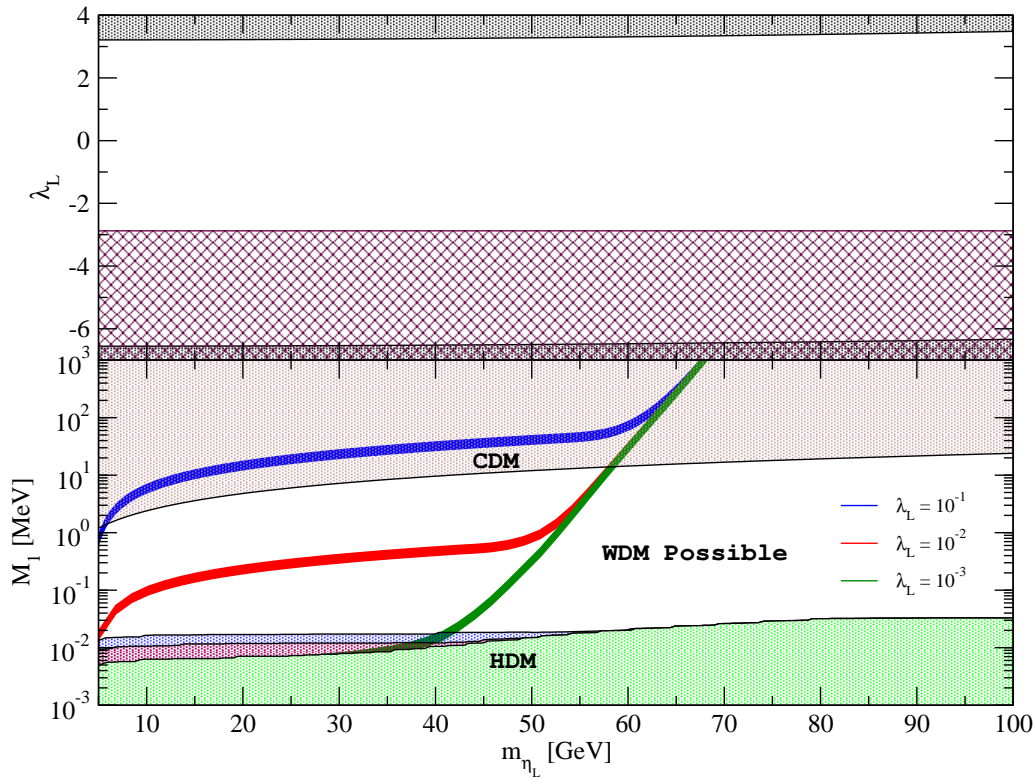} 
\caption{Same as in Fig.~\ref{relicplot-1} but for $M_{\rm Higgs} =
  500$ GeV, $m_{\eta_H} = 150$ GeV and $m_{\eta^\pm} = 300$ GeV.}  
\label{relicplot-2}
\end{figure}

In Figs.~\ref{relicplot-1} and~~\ref{relicplot-2}, we show regions of
the $m_{\eta_L} -M_1$ plane in which $N_1$ has the right dark matter
density for two different sets of parameters. The Higgs mass is $M_{\rm
  Higgs}=$~160~GeV, $m_{\eta_H} = 125$ GeV and $m_{\eta^\pm} =
130$~GeV in Fig.~\ref{relicplot-1} and the Higgs mass is $M_{\rm
  Higgs}=$~500 GeV, $m_{\eta_H} = 150$ GeV and $m_{\eta^\pm} =
300$~GeV in Fig.~\ref{relicplot-2}. The upper panels of the figures
show the bounds on $\lambda_L$ obtained from vaccum stability 
(cross-hatched violet regions) and pertubativity (shaded gray region) 
arguments. From top to bottom, the blue, red and green colored narrow
bands in the lower panels of Figs.~\ref{relicplot-1}
and~\ref{relicplot-2} show the regions in the $m_{\eta_L}$- $M_1$
plane in which $\Omega_{N_1} h^2$ in Eq.~\ref{Omega} is within the
$3\sigma$ measured range for the dark matter (either CDM or WDM). The
different colors of the narrow bands indicate different values of
$\lambda_L$, as shown in the panels. The unshaded background region
labeled ``WDM Possible" corresponds to the range in Eq.~\ref{m-WDM},
where the $N_1$ may constitute WDM (above it, it can only be CDM and
below it, only  HDM). For any particular value of $T_{\rm{decay}}$
between 5~MeV (which defines the upper boundary of the unshaded region)
and $m_{\eta_L}/x_f$ (which defines the lower boundary of the unshaded
region) there is one value of $M_1$ given by Eq.~\ref{m-WDM}, within
the unshaded background region for which $N_1$ would be WDM ($N_1$
would be CDM for all larger values of $M_1$ and HDM for all smaller
ones). Notice that the lower boundary of the unshaded region depends
on $\lambda_L$ through $x_f$, thus the blue, red, green colors of the
lower regions, for which the $N_1$ can only be WDM. Thus, within the
unshaded background region $N_1$ could be WDM or CDM, depending on
$T_{\rm decay}$. For a given set of parameters defining the model (and
hence a given $T_{\rm{decay}}$), in order for $N_1$ to be allowed as
HDM, its mass $M_1$ must be, at least, a factor of $\Omega_{\rm{DM}} h^2
/\Omega_{\rm{HDM-max}} h^2 = 0.1099/0.014 \simeq 8$ smaller than the value
at center of the colored band defined by Eq.~\ref{Omega} for a given
$m_{\eta_L}$. The figures show that the $N_1$ could be HDM  even for
masses as large as $\sim$~1~keV.

In conclusion, we have shown that Inert-Sterile neutrinos, produced
non-thermally in the early Universe, could be a viable WDM or CDM
candidate. They are virtually non-detectable in either direct or
indirect dark matter searches because of their extremely weak
couplings to SM particles. Thus, their existence could be revealed only
by discovering other particles of the model in collider
experiments. We should keep in mind that the dark matter may consist
of an admixture of different types of particles and that particles 
undetectable in dark matter searches may be part of it. The existence
of these particles could only be inferred from collider data,
supplemented by the null results from dark matter searches or with
results from these searches which find other detectable dark matter
components with a density smaller than required to constitute the
whole of the dark matter. Unveiling the nature of the dark 
matter does necessarily require the combination of collider and direct
and indirect searches.

\vspace{0.5cm}

{\bf Acknowledgements}\\

We thank E.~Dolle and S.~Su for helpful discussions. This work was
supported in part by the US Department of Energy Grant
DE-FG03-91ER40662, Task C at UCLA. SPR is partially supported by the 
Portuguese FCT through CERN/FP/83503/2008 and CFTP-FCT UNIT 777, which
are partially funded through POCTI (FEDER), and by the Spanish Grant
FPA2008-02878 of the MCT. GG and SPR would like to thank the Aspen
Center for Physics and CERN where part of this work took place. EO
would also like to thank CERN for hospitality.

\end{document}